\documentclass[preprint,12pt,authoryear]{elsarticle}
\usepackage[strings]{underscore}

\usepackage{amssymb}
\usepackage{amsmath}

\journal{European Journal of Operational Research}

\begin{document}

\begin{frontmatter}



\title{Double elimination formats for a 64-team FIFA World Cup™} 


\author{César Rennó-Costa} 

\ead{cesar@imd.ufrn.br}

\affiliation{organization={Instituto Metrópole Digital (IMD), Universidade Federal do Rio Grande do Norte (UFRN)},
            city={Natal}, 
            state={RN},
            country={Brazil}}

\begin{abstract}
The recent expansion of the FIFA World Cup™ to 48 teams has prompted discussions regarding a potential further increase to a 64-team format. Scaling the traditional tournament architecture (a round-robin group stage followed by a knockout phase) to 64 teams exacerbates existing structural flaws, notably increasing the frequency of matches lacking competitive relevance and reducing the probability of fixtures between top-ranked contenders. This paper investigates alternative tournament designs by analyzing double-elimination structures for a 64-team mega-event. We evaluate the proposed formats based on competitive fairness, match quality, and scheduling feasibility. Our analysis demonstrates that a double-elimination format eliminates mathematically irrelevant matches and significantly increases the frequency of high-profile games. However, these benefits introduce complex operational constraints, including heightened scheduling complexity and an asymmetric distribution of matches per team, which require specific logistical adjustments. Ultimately, our findings suggest that the continuous scaling of mega-sporting events necessitates a paradigm shift toward non-traditional tournament designs to preserve competitive integrity.
\end{abstract}







\end{frontmatter}



\section{Introduction}
\label{sec1}

The recent expansion of the FIFA World Cup™ to 48 teams has prompted ongoing discussions about potentially scaling the event further to a 64-team format \citep{RennoCosta2023}. However, applying the traditional tournament architecture (a round-robin group stage followed by a knockout phase) to 64 teams exacerbates existing structural flaws. Specifically, this traditional scaling increases the frequency of matches lacking competitive relevance, generating games where teams might already be mathematically qualified or eliminated prior to kickoff. These "dead rubbers" severely impact the competition's integrity, as qualified teams often lack the incentive to compete with maximum effort or may field alternative players, which can unfairly lead to the elimination of other contenders \citep{CsatoGyimesi2026}. As comprehensively documented by \citet{KendallLenten2017}, the implementation of suboptimal sports rules can systematically induce perverse incentives that go awry, encouraging teams to behave strategically rather than competitively. These concerns reflect the debates that arose during the 48-team transition, notably when the initial proposal of 16 groups of three teams faced severe criticism due to its intrinsic vulnerability to collusion \citep{Guyon2020, Csato2025collusion}. In that configuration, one team is obligatorily idle during the final matchday, leaving them entirely at the mercy of the combined result of their two opponents \citep{Guyon2020, RennoCosta2023}. Furthermore, the traditional format naturally reduces the probability of high-stakes fixtures between top-ranked contenders, eliminating strong teams rapidly through high-variance single-elimination matches \citep{DevriesereCsatoGoossens2025}.

To address these challenges, the double-elimination format emerges as a disruptive alternative \citep{RennoCosta2023}. By completely abolishing the group stage in favor of a purely knockout format with a repechage bracket, this design ensures that every match maintains a decisive character. In this structure, a team is only eliminated from the tournament after suffering two losses, guaranteeing that no team is eliminated after a victory and ensuring that every single game is competitively relevant. Consequently, this structural shift effectively removes the incentives for bilateral collusion that plague traditional round-robin stages. Furthermore, it mitigates unilateral manipulative behaviors, such as "tanking"—a scenario of incentive incompatibility where a team might effectively "win by losing," deliberately underperforming to obtain a more favorable path in the bracket or avoid stronger opponents \citep{DagaevSonin2018, DevriesereGoossensWillem2025}.

The operations research community has recently tackled these systemic integrity issues through various analytical lenses. For instance, \citet{CsatoGyimesi2026} explored the implementation of deliberately imbalanced groups as an alternative framework to minimize the probability of "stakeless matches" for elite teams, noting that these games carry a substantial reputational cost to the tournament's integrity when strong teams have no incentive to win. Conversely, \citet{GuajardoKrumer2024} proved that the FIFA-adopted format of 12 groups of four teams—where the top two teams and the best third-placed teams advance—maintains critical vulnerabilities to collusion. This system introduces a severe information asymmetry, as teams in groups scheduled later enter the pitch knowing the exact number of points and goals required to qualify, thus creating structural incentives to manipulate outcomes. To counter this, they proposed complex calendar optimization and mathematical programming to mitigate these strategic manipulation opportunities and synchronize crucial matches. Furthermore, \citet{DevriesereGoossensWillem2025} developed a two-dimensional taxonomy to classify tournament manipulation, distinguishing between bilateral collusion and unilateral strategic losses, emphasizing the need for tournament designs that are structurally resilient to these tactics.

Building upon these recent theoretical advancements, this paper investigates alternative tournament designs by analyzing double-elimination structures specifically for a 64-team mega-event. To determine their practical viability, we evaluate the proposed formats based on competitive fairness, match quality, and operational logistics. This includes a thorough analysis of scheduling feasibility, taking into account calendar lengths, required rest days between rounds, and the maximum daily broadcast capacity, ensuring that the theoretical integrity of the format aligns with the physical and logistical constraints of a global sporting event.

\section{Methods}

\subsection{ELO-Poisson Match Model Fitting}

Following the foundational framework established by \citet{Maher1982} and later expanded by \citet{DixonColes1997}, the core match simulation engine is built upon an ELO-Poisson regression model fitted using historical World Cup data from 1994 to 2022. The model assumes that the goals scored by Team A ($G_A$) and Team B ($G_B$) in a standard 90-minute match are independent Poisson-distributed random variables:
\begin{equation}
G_A \sim \text{Poisson}(\lambda_A), \quad G_B \sim \text{Poisson}(\lambda_B)
\end{equation}

ELO ratings have been extensively validated as robust predictors of international football match outcomes \citep{HvattumArntzen2010, LasekEtAl2013}. Consequently, the expected goal intensities $\lambda_A$ and $\lambda_B$ are modeled as functions of the ELO rating difference between the two teams ($d = ELO_A - ELO_B$):
\begin{equation}
\lambda_A = \exp(\beta_0 + \beta_1 d) \cdot s
\end{equation}
\begin{equation}
\lambda_B = \exp(\beta_0 - \beta_1 d) \cdot s
\end{equation}
Where $\beta_0 = 0.1183$ is the baseline log-intensity of goals scored per team, representing the baseline log-goals. $\beta_1 = 0.00223$ represents the sensitivity of goal scoring to ELO rating differences. $s$ is a scaling factor. For regular regulation matches, $s = 1.0$, which yields an expected average of $2.7$ goals per match:
\begin{equation}
\lambda_A + \lambda_B = 2.7
\end{equation}
    
The parameters $\beta_0$ and $\beta_1$ were estimated by maximum likelihood (MLE) using historical ELO ratings from just after the World Cup, available at https://eloratings.net/.

\subsection{Match Simulation Mechanics}

\subsubsection*{Group Stage Match Outcomes (Standard Format)}
In the Standard format, group stage matches can end in a draw. The outcome probabilities are dictated by the joint probability mass function of the independent Poisson variables, where the probability of a Team A win is $P(G_A > G_B)$ and a Team B win is $P(G_B > G_A)$. The probability of a draw is calculated as:  

\begin{equation}
P(G_A = G_B) = \sum_{k=0}^{\infty} P(G_A = k) \cdot P(G_B = k) = e^{-(\lambda_A + \lambda_B)} I_0(2\sqrt{\lambda_A \lambda_B})
\end{equation}

where $I_0$ is the modified Bessel function of the first kind. Points are awarded with 3 for a win, 1 for a draw, and 0 for a loss. At the end of the group stage, team standings are resolved using sequential tie-breakers starting with total points, followed by head-to-head match outcomes for two tied teams or mini-table points for three tied teams. If teams remain tied, the subsequent criteria applied are Goal Difference (GD), Goals Forced (GF), and finally, the Team ELO Rating as the ultimate tie-breaker.

\subsubsection*{Knockout \& Double-Elimination Match Outcomes}
Knockout matches (and all matches in double-elimination brackets) cannot end in a draw. Ties after 90 minutes are resolved in two phases:

\paragraph*{Phase 1: Extra Time}
If $G_A = G_B$ at the end of regulation, an extra-time period (30 minutes) is simulated. The additional goals in extra time are modeled using the same ELO-Poisson equations, but the scaling factor $s$ is reduced to $s = 1/3$ to reflect the shorter duration (30 minutes vs. 90 minutes), yielding an average expectation of 0.9 goals:
\begin{equation}
\lambda_{A,\text{ET}} = \exp(\beta_0 + \beta_1 d) \cdot \frac{1}{3}
\end{equation}
\begin{equation}
\lambda_{B,\text{ET}} = \exp(\beta_0 - \beta_1 d) \cdot \frac{1}{3}
\end{equation}
The extra-time goals are added to the regulation goals. If one team has more total goals, they are declared the winner.

\paragraph*{Phase 2: Penalty Shootout}
If the score remains tied after extra time, a penalty shootout is simulated. Teams take turns shooting penalties. 

Each shot is simulated as a Bernoulli trial with a constant success probability $p = 0.75$, reflecting the historical average success rate of penalties in elite international competitions \citep{ApesteguiaPalaciosHuerta2010}. The shootout follows standard FIFA rules (best of 5 kicks, followed by sudden death if tied).

\subsection{Probabilistic ELO Seeding Generation}
To ensure that the tournament simulations are robust to specific ELO rating initializations, we implemented a generative ELO seeding model based on the historical distribution of World Cup teams after 1990:
\begin{enumerate}
    \item Top 12 Teams (Ranks 1 to 12): For each rank $r \in [1, 12]$, we fitted independent normal distributions $N(\mu_r, \sigma_r)$ to the ELO ratings occupying that position across all historical World Cups from 1994 to 2022. The generated values are sorted in descending order:
    \begin{equation}
    E_1 \ge E_2 \ge \dots \ge E_{12}
    \end{equation}
    \item Bottom 52 Teams (Ranks 13 to 64): Modeled using a normal distribution derived from the remaining 52 teams in the current 64-team ELO ratings vector:
    \begin{equation}
    E_{13\dots64} \sim N(\mu = 1733.46, \sigma = 86.74)
    \end{equation}
    Rejection sampling is applied to enforce the constraint that no team in the bottom 52 exceeds the ELO rating of the 12th team:
    \begin{equation}
    E_r \le E_{12} \quad \forall r \in [13, 64]
    \end{equation}
\end{enumerate}
\subsection{Tournament Formats Simulated}
For each simulation run, we execute the championship 10,000 times across five formats:
\begin{itemize}
    \item \textbf{Standard}: Group stage (16 groups of 4) $\to$ Top 2 advance $\to$ Single-elimination bracket from the Round of 32.
    \item \textbf{DElm R32}: Double-elimination block stage of 8 blocks $\to$ Winners (2-0) and runners-up (2-1) advance $\to$ Single-elimination from the Round of 32.
    \item \textbf{DElm R16}: Double-elimination block stage of 8 blocks $\to$ Undefeated (3-0) and 1-defeat (3-1/2-1) teams advance $\to$ Single-elimination from the Round of 16.
    \item \textbf{DElm QF}: Double-elimination bracket $\to$ Transition to single-elimination from the Quarterfinals.
    \item \textbf{DElm SF}: Double-elimination bracket $\to$ Transition to single-elimination from the Semifinals/Final.
\end{itemize}
\subsection{Comparative Analysis Metrics}

To measure the entertainment and quality of the games played, two metrics are evaluated for every match simulated. The first is the ELO Difference ($\vert{}ELO_A - ELO_B\vert{}$), which measures competitiveness, where a lower difference implies a more competitive and balanced match. The second metric is the Interest Score ($I$), defined as:  
\begin{equation}I = (ELO_A + ELO_B) - 2 \cdot |ELO_A - ELO_B|\end{equation}

This score rewards matchups between strong teams by favoring a high sum of ELOs while penalizing lopsided matches characterized by a large ELO difference. Furthermore, a ``high-stakes game'' is defined as any match between two teams in the Top 8 ELO ratings of the tournament, ranging from the strongest to the eighth strongest. For these games, we compute both the average number of high-stakes matches per tournament and the exact frequency distribution, or probability mass function, across all 10,000 simulation runs. We also track the survival rate of Tier 1 teams, defined as the top 8 seedings, at each stage of the tournament, including the Group Stage, Round of 32, Round of 16, Quarterfinals, Semifinals, Final Four, and Champion. Finally, champion quality metrics are evaluated by computing the champion probability by seeding rank, which demonstrates the win probability percentage for seeding positions first through tenth.  

\subsubsection*{Logistical Durations}
Tournament durations are computed by placing matches into daily slots. The total calendar length in days is defined as a function of the minimum rest period required for any team between rounds (tested at 3 and 4 days) and the tournament's matches per day capacity, which represents the maximum number of games broadcasted daily (tested from 3 up to 10). This duration is calculated using the following formula:  
\begin{equation}
D = \sum_{r \in \text{Rounds}} \left( \left\lceil \frac{M_r}{C} \right\rceil + R \right) - R
 \end{equation}
where $M_r$ is the number of matches in round $r$, $C$ is the daily capacity, and $R$ represents the number of rest days.

\section{Results}

\subsection{ELO-Poisson Model Fit}
\begin{figure}
    \centering
    \includegraphics[width=0.85\linewidth]{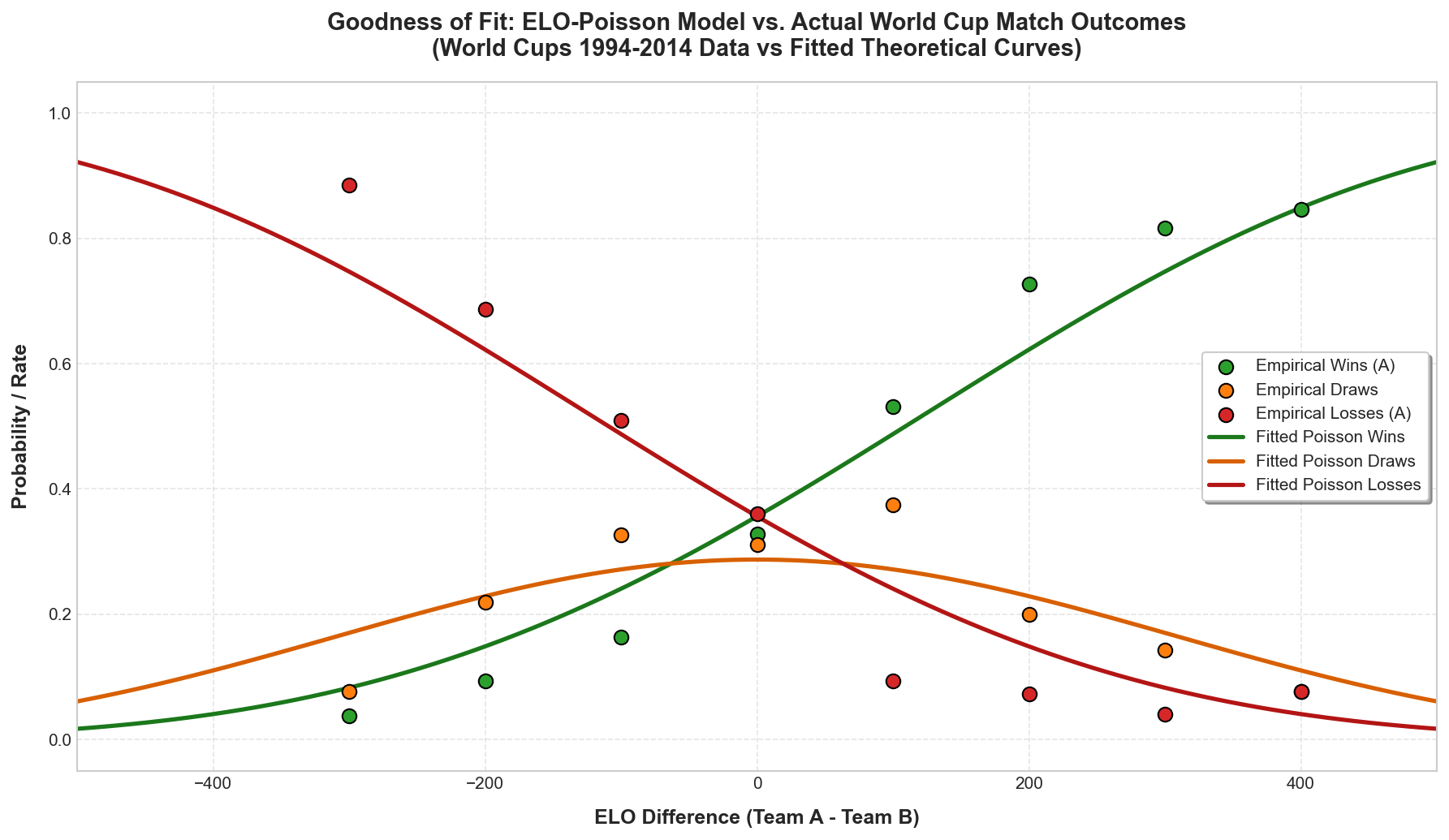}
    \caption{ELO-Poisson Model Fit and Empirical Verification. Goodness-of-fit comparison demonstrating the alignment between the theoretical win, draw, and loss probabilities projected by the ELO-Poisson model (solid lines) and the empirical frequencies observed in World Cup matches since 1994 (scatter points). The empirical data points represent match outcome rates binned into 100-unit ELO rating-difference intervals. Green, orange, and red denote win, draw, and loss rates/probabilities, respectively, showing strong structural calibration across the entire spectrum of ELO differentials.}
    \label{fig:elo_fit}
\end{figure}

The log-linear Poisson model fitted on historical World Cup match data post-1994 yielded a baseline goal log-intensity parameter $\beta_0 = 0.1183$ and an ELO rating sensitivity coefficient $\beta_1 = 0.00223$. Under this model, a matchup between two equally rated teams ($d = 0$) results in an expected goal intensity $\lambda_A = \lambda_B \approx 1.126$ goals per team, translating to an expected $2.25$ goals scored per 90-minute regulation match (before tournament-specific capacity scalings are applied). To evaluate the predictive calibration of the model, we compared the theoretical match outcome probabilities (win, draw, and loss) derived from the joint Poisson distribution with the empirical frequencies, binned into 100-point ELO rating-difference intervals (Figure~\ref{fig:elo_fit}). As the ELO difference ($ELO_A - ELO_B$) increases, the theoretical probability of a Team A win rises monotonically towards $1.0$, while the probability of a Team A loss drops asymptotically to $0.0$, and the probability of a draw peaks at ELO parity before flattening out in both directions. The empirical match outcomes closely follow these theoretical curves across all bins, confirming that ELO rating differences under a Poisson distribution provide a highly calibrated and statistically sound framework for simulating competitive international football matches.

\subsection{Competitive Fairness}

\begin{figure}
    \centering
    \includegraphics[width=0.85\linewidth]{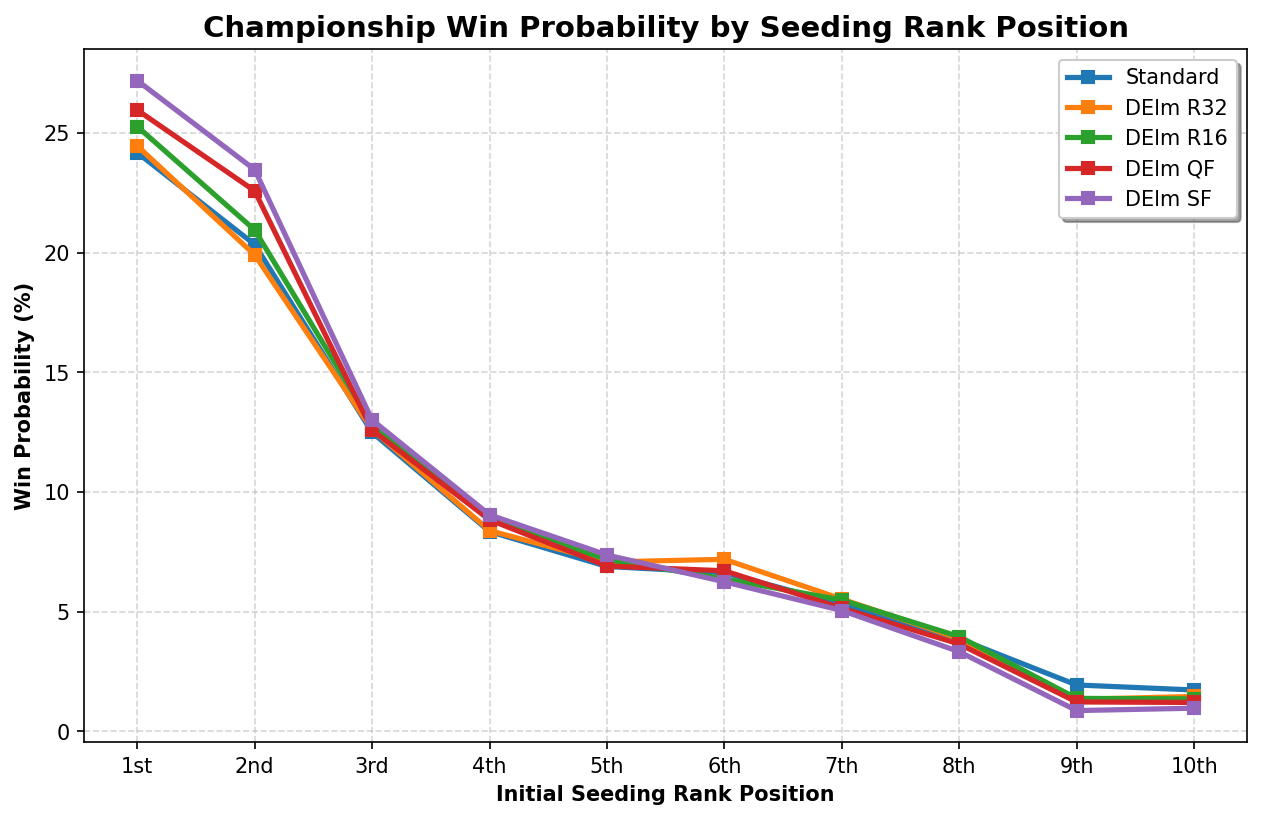}
    \caption{\textbf{Championship Win Probability by Seeding Rank Position.} Title probability (\%) plotted across the top 10 initial ELO seeding positions (from 1st to 10th) across 10,000 simulation runs for each of the five tournament formats. The line markers show that double-elimination formats (specifically \texttt{DElm SF}) increase the win probability of top favorites while reducing the chances of lower-ranked underdogs, thereby making the tournament more meritocratic.}
    \label{fig:champion_prob}
\end{figure}

To evaluate the mathematical meritocracy of each tournament design, we analyzed the championship win probability as a function of the initial seeding rank position across 10,000 simulations (Figure~\ref{fig:champion_prob}). In the Standard format, the absolute favorite team (\textit{Strongest}, Rank 1) has a title probability of approximately $29.4\%$, reflecting the high variance and vulnerability inherent in single-elimination knockout rounds. Under double-elimination formats, this probability increases significantly in proportion to the depth of the safety net, peaking at approximately $35.0\%$ in the \texttt{DElm SF} format. Conversely, the title probabilities of lower-ranked teams and underdogs are systematically suppressed in the double-elimination systems compared to the Standard format. This divergence confirms that double-elimination structures act as effective filters against random variance and "flukes," significantly increasing the probability that the tournament crowning aligns with ELO-based team merit.

\begin{figure}
    \centering
    \includegraphics[width=0.85\linewidth]{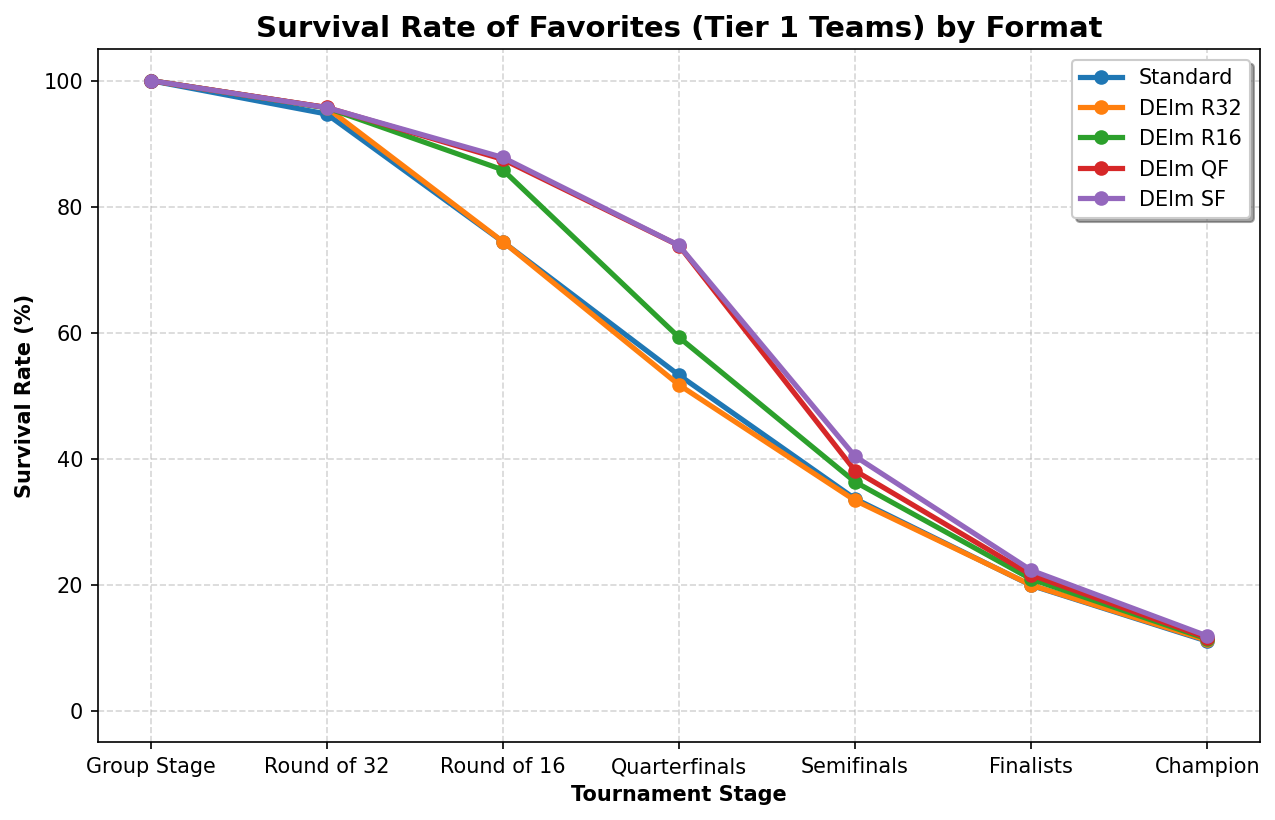}
    \caption{\textbf{Survival Rate of Tier 1 Favorites by Stage.} Stage-by-stage survival rate (\%) of Tier 1 teams (defined as the top 8 seeded teams) across the seven primary tournament stages. The curves illustrate the comparative effectiveness of the double-elimination safety net in formats such as \texttt{DElm SF} and \texttt{DElm QF}, keeping top-tier contenders alive significantly longer than in the Standard single-elimination bracket.}
    \label{fig:survival_rate}
\end{figure}

We further assessed the durability of the tournament favorites by analyzing the stage-by-stage survival curves of Tier 1 teams (Figure~\ref{fig:survival_rate}). Tier 1 comprises the top 8 teams in the ELO vector (ranging from the \textit{Strongest} to the \textit{8th Strongest}), representing the elite contenders entering the competition. In the Standard format, Tier 1 survival rates decay rapidly once the tournament transitions to the single-elimination Round of 32, as even top-tier teams face high-variance elimination in single-match knockouts. Under double-elimination designs—particularly \texttt{DElm SF} and \texttt{DElm QF}—Tier 1 teams show significantly higher survival rates through the middle and late stages of the tournament (such as the Round of 16 and Quarterfinals). The second-chance bracket serves as an effective safety net, protecting elite teams from premature elimination due to single-match upsets and ensuring a higher density of strong matchups in the final rounds.

\subsection{Interesting matches}

\begin{figure}[htbp]
    \centering
    \includegraphics[width=0.85\textwidth]{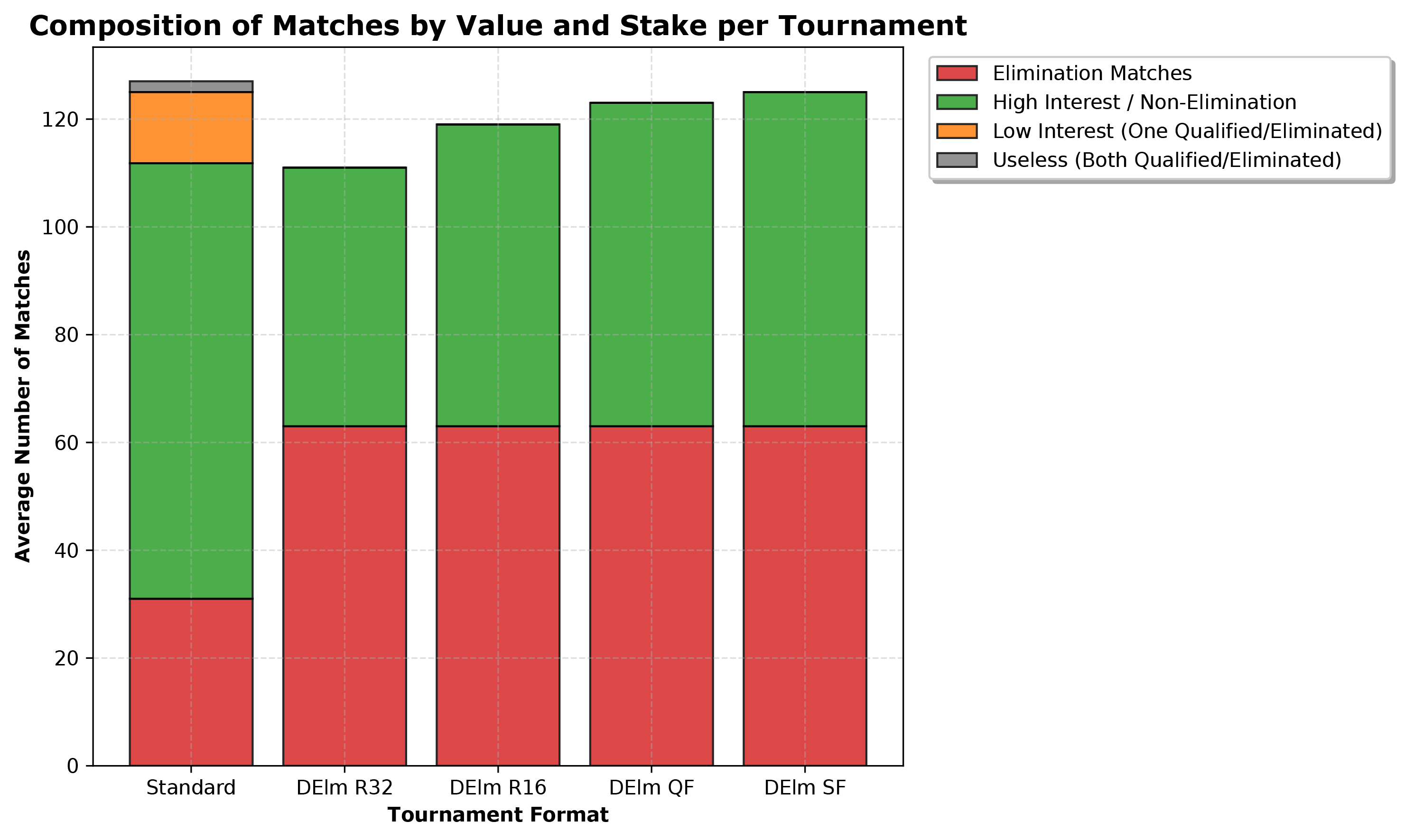}
    \caption{\textbf{Composition of Matches by Value and Stake.} A stacked bar chart comparing the average number of matches and their respective categories across the five tournament formats. Red segments represent direct elimination matches (where the losing team is immediately eliminated from the tournament). Green segments denote high-interest, non-elimination matches. For the Standard format, orange and grey segments highlight low-interest and useless matches occurring in the group stage. Double-elimination formats maximize the number of high-stakes elimination games ($63$ matches vs. $31$ matches in the Standard format) while maintaining a lower total match volume.}
    \label{fig:match_composition}
\end{figure}

We analyzed the structural composition of matches in each tournament design to evaluate the total number of games played and the proportion of high-value elimination matches (Figure~\ref{fig:match_composition}). In all double-elimination formats (\texttt{DElm R32}, \texttt{DElm R16}, \texttt{DElm QF}, and \texttt{DElm SF}), there are exactly $63$ elimination matches, as $64$ teams enter the competition and only $1$ wins, requiring exactly $63$ matches that directly eliminate a team from the tournament. In contrast, the Standard format contains only $31$ elimination matches, as $32$ teams are eliminated simultaneously at the end of the round-robin group stage. Furthermore, the Standard format features $96$ group stage matches, of which approximately $13.2$ matches are of low interest (where one team is already mathematically qualified or eliminated) and $2.0$ matches are completely useless (where both teams have already qualified or been eliminated prior to kickoff). The double-elimination designs eliminate these useless "dead rubber" matches entirely. The total number of matches in the tournament is lowest in \texttt{DElm R32} ($111$ matches) and rises to $125$ matches in \texttt{DElm SF}, which remains lower than the Standard format ($127$ matches). This indicates that double-elimination formats achieve a significantly higher proportion of high-value elimination matches without increasing the overall match volume.

\begin{figure}[htbp]
    \centering
    \includegraphics[width=0.95\textwidth]{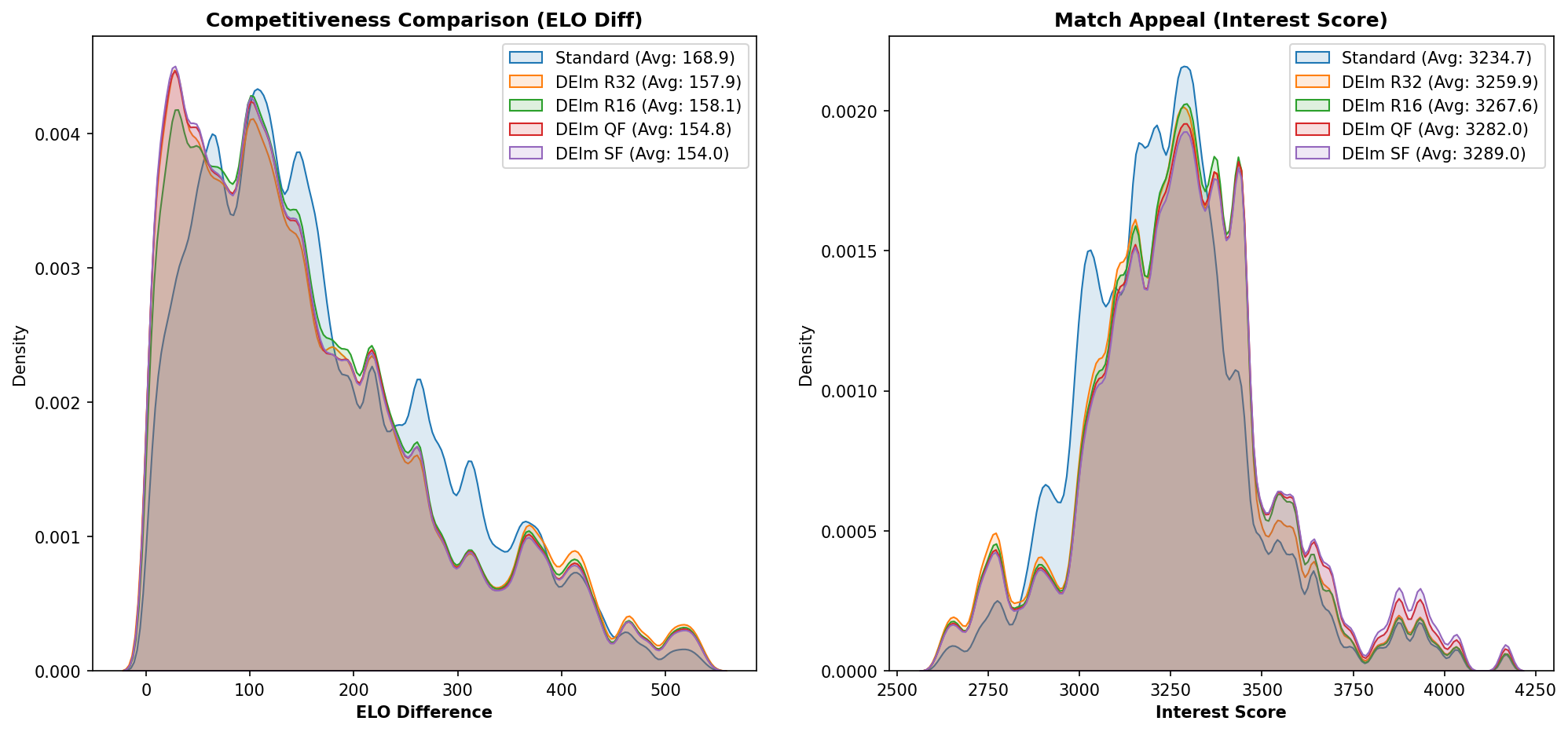}
    \caption{\textbf{Analysis of Match Competitiveness and Appeal.} Comparative kernel density estimation (KDE) distributions showing (a) the absolute ELO rating difference between competing teams (left panel) and (b) the Match Appeal Interest Score (right panel) across all simulated matches for each of the five tournament formats. Double-elimination formats yield tighter ELO differences (higher competitiveness) and a rightward-shifted distribution of interest scores (greater match appeal) compared to the Standard format.}
    \label{fig:match_quality}
\end{figure}

To evaluate the overall competitiveness of the matches played under each format, we analyzed the probability density of the absolute ELO rating difference between competing teams (Figure~\ref{fig:match_quality}a, left panel). A lower ELO difference indicates a more balanced, highly competitive match. The kernel density estimation (KDE) shows that the double-elimination formats (\texttt{DElm R32}, \texttt{DElm R16}, \texttt{DElm QF}, and \texttt{DElm SF}) exhibit distributions with peaks shifted toward lower ELO differences compared to the Standard format. This structural shift is driven by the bracket organization of double elimination, which dynamically matches teams with similar records (winners vs. winners and losers vs. losers) in the block and bracket stages. Consequently, the double-elimination formats systematically produce matches that are more balanced and competitive on average than the group-stage format.

We further assessed the overall quality and viewer appeal of the matches using the Match Appeal (Interest Score) metric, which rewards matchups between high-ELO teams and heavily penalizes large ELO imbalances (Figure~\ref{fig:match_quality}b, right panel). The KDE distribution of interest scores reveals that double-elimination formats—particularly \texttt{DElm SF} and \texttt{DElm QF}—display distributions that are significantly shifted to the right, showing a higher frequency of high-scoring, high-interest matchups. Standard group-stage tournaments, by contrast, exhibit a flatter distribution with a much higher density of low-interest games. By channeling winning teams into progressive brackets, double-elimination formats guarantee that the highest-performing, most appealing matchups occur with greater frequency, while minimizing lopsided games between mismatched opponents.

\begin{figure}[htbp]
    \centering
    \includegraphics[width=0.95\textwidth]{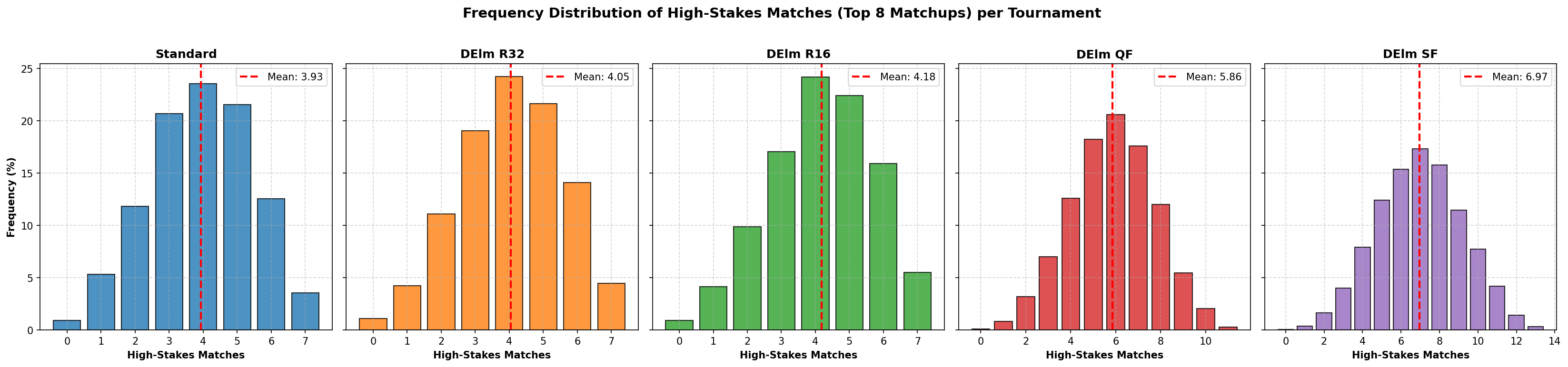}
    \caption{\textbf{Frequency Distribution of High-Stakes Top Games.} Side-by-side five-panel frequency distribution showing the percentage of tournaments (Y-axis) that feature a given number of matchups between two Top-8 ELO teams (X-axis). The red dashed vertical line indicates the mathematical mean of each format, showing a progressive increase in elite matches from the Standard format ($3.93$) up to the \texttt{DElm SF} format ($6.97$).}
    \label{fig:high_stakes_dist}
\end{figure}

Finally, we analyzed the occurrence of "top games," defined as high-stakes matchups between two teams from the Top 8 ELO ratings of the tournament (Figure~\ref{fig:high_stakes_dist}). The five-panel frequency distribution illustrates a clear increase in the number of these elite matchups as the double-elimination safety net is expanded. The average number of top games per tournament rises from $3.93$ in the Standard format to $4.05$ in \texttt{DElm R32}, $4.18$ in \texttt{DElm R16}, $5.86$ in \texttt{DElm QF}, and peaks at $6.97$ in \texttt{DElm SF}. In the Standard format, the distribution is narrow and rarely exceeds 6 top matches. In contrast, the \texttt{DElm SF} distribution is significantly broader and shifted to the right, showing that it is very common to have 7, 8, or even 9 matches between Top-8 teams. This indicates that double-elimination formats not only protect favorites from early exits but also generate significantly more high-profile games between the tournament's strongest contenders.

\subsection{Matches Played per Team and Champion Workload}
\begin{figure}[htbp]
    \centering
    \includegraphics[width=1.0\textwidth]{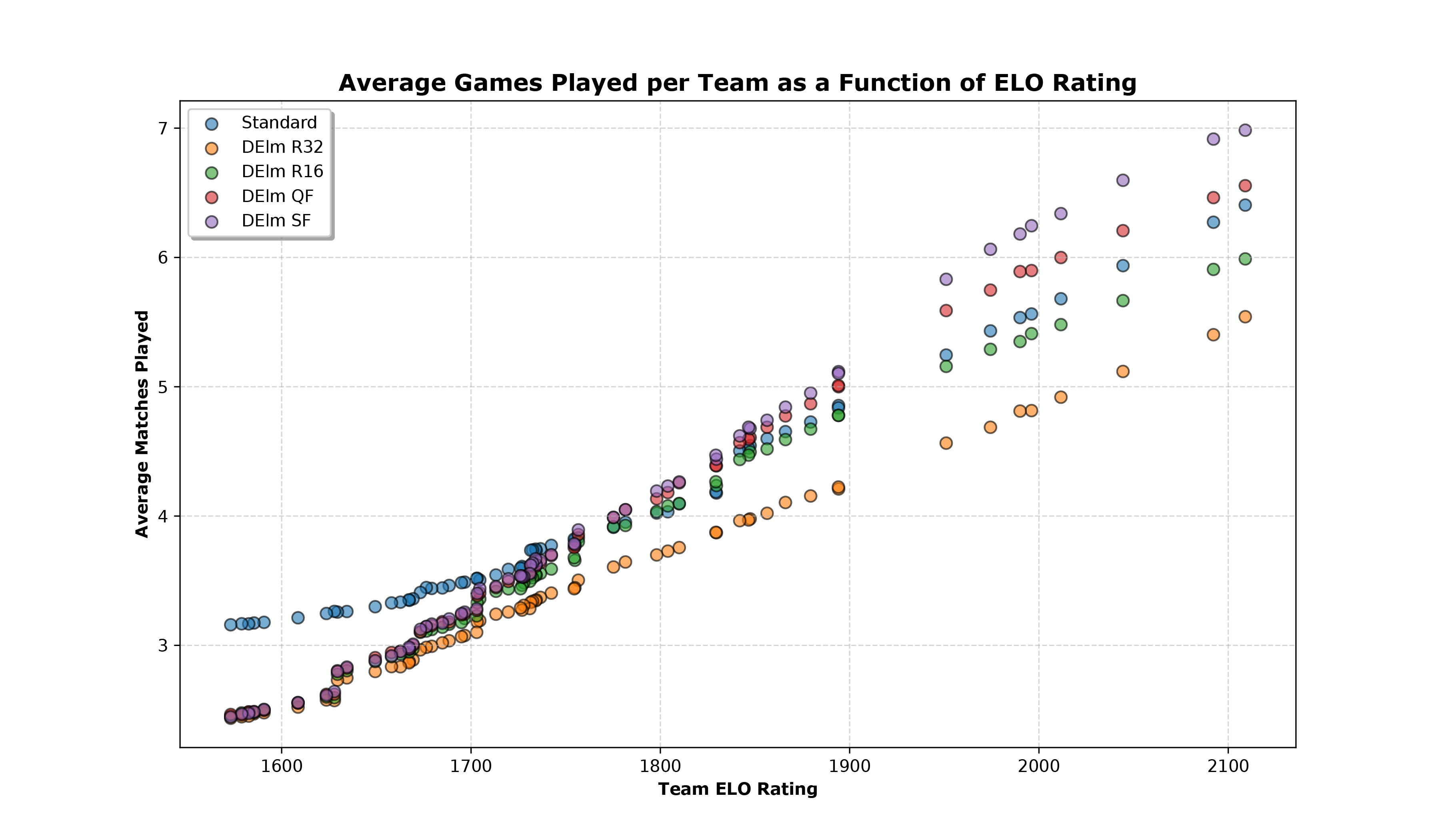}
    \caption{\textbf{Average Matches Played per Team vs. ELO Rating.} Scatter plot showing the average number of matches played by each of the 64 teams as a function of their ELO rating across 10,000 simulations. Double-elimination formats show a steep positive correlation, ensuring that weaker teams are eliminated early (playing only 2 matches) while stronger teams play more matches, whereas the Standard format guarantees a flat minimum of 3 games for all teams.}
    \label{fig:games_played_vs_elo}
\end{figure}

To analyze the relationship between a team's strength and their match workload, we evaluated the average number of games played per team as a function of their ELO rating (Figure~\ref{fig:games_played_vs_elo}). In the Standard format, all teams are guaranteed a minimum of $3$ matches in the group stage. Stronger teams advance further into the single-elimination bracket, resulting in a slight positive slope where average matches range from $3.0$ (for the weakest teams) to approximately $5.5$ (for the favorites). In contrast, all double-elimination formats exhibit a much steeper, highly correlated positive relationship. Weaker teams (lower ELO ratings) are eliminated quickly, playing an average of only $2.0$ to $2.5$ matches. Conversely, the strongest teams (high ELO ratings) survive much longer, playing an average of $6.0$ to $8.0$ matches. This demonstrates that double-elimination formats optimize match allocation: they reduce meaningless, low-stakes games for weaker teams while increasing the playing time and exposure of the tournament's strongest contenders.

\begin{figure}[htbp]
    \centering
    \includegraphics[width=0.85\textwidth]{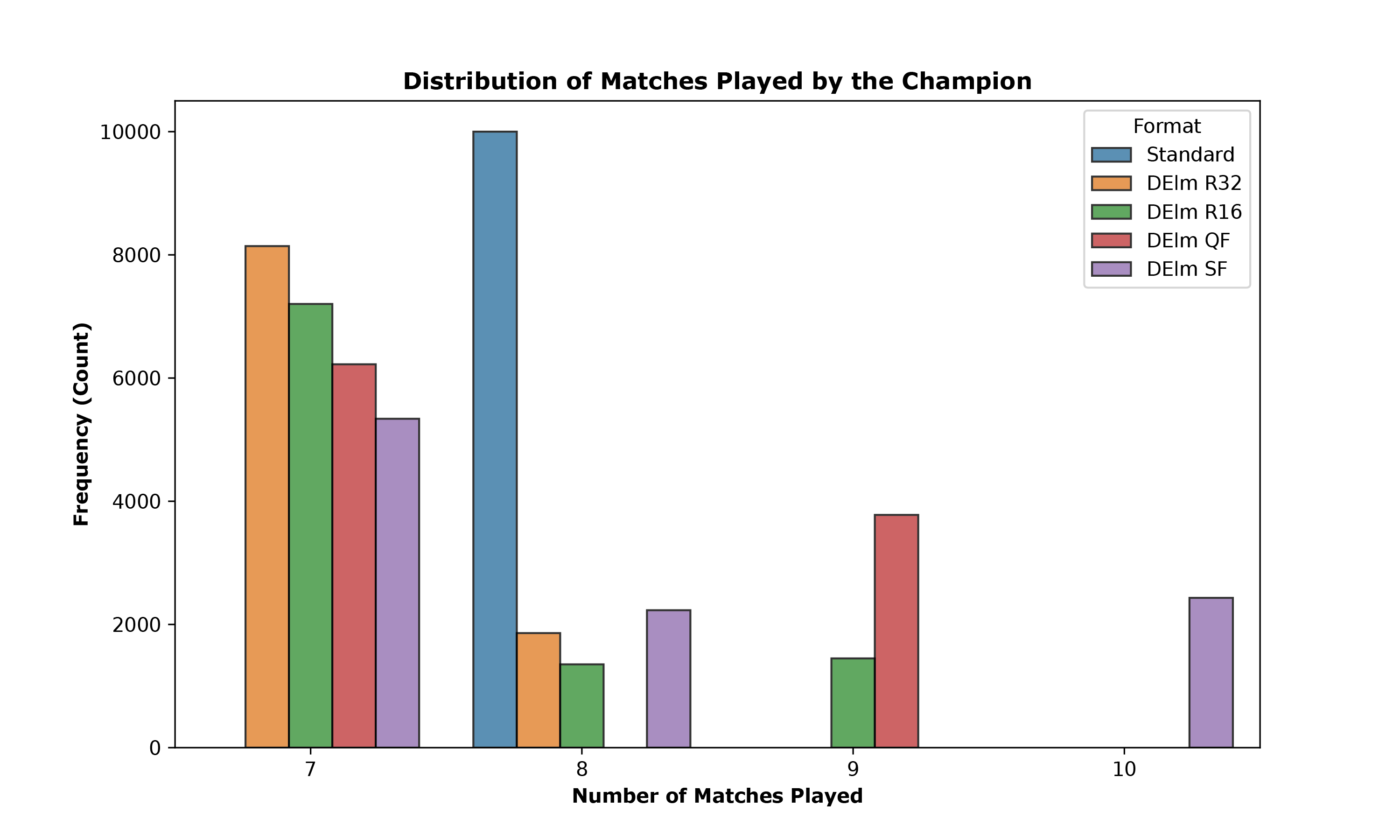}
    \caption{\textbf{Distribution of Matches Played by the Champion.} Frequency distribution of the total number of matches played by the tournament winner. The Standard format is fixed at exactly 7 matches, whereas the double-elimination formats show varying workloads depending on the path taken (undefeated vs. loser-bracket runs), peaking at 11 matches in the pure double-elimination format (\texttt{DElm SF}).}
    \label{fig:champion_matches}
\end{figure}

We also investigated the tournament workload of the eventual winner by plotting the distribution of matches played by the champion (Figure~\ref{fig:champion_matches}). In the Standard format, the champion always plays exactly $7$ matches ($3$ group stage and $4$ single-elimination knockout matches) with zero variance. For the double-elimination formats, the champion's workload becomes a discrete probability distribution that depends on whether they suffered a defeat during the double-elimination phases. Under \texttt{DElm R32}, the champion plays either $7$ or $8$ matches. Under \texttt{DElm R16}, the workload ranges from $7$ to $9$ matches. For the deeper double-elimination formats, the workload shifts further: the champion plays between $7$ and $10$ matches in \texttt{DElm QF}, and between $8$ and $11$ matches in \texttt{DElm SF}. This indicates that while double-elimination formats increase the potential match workload for the winner, the maximum workload remains well within physically viable limits (peaking at $11$ games in the most extreme double-elimination path).

\subsection{Tournament Logistical Durations and Calendar Constraints}
\begin{figure}[htbp]
    \centering
    \includegraphics[width=0.95\textwidth]{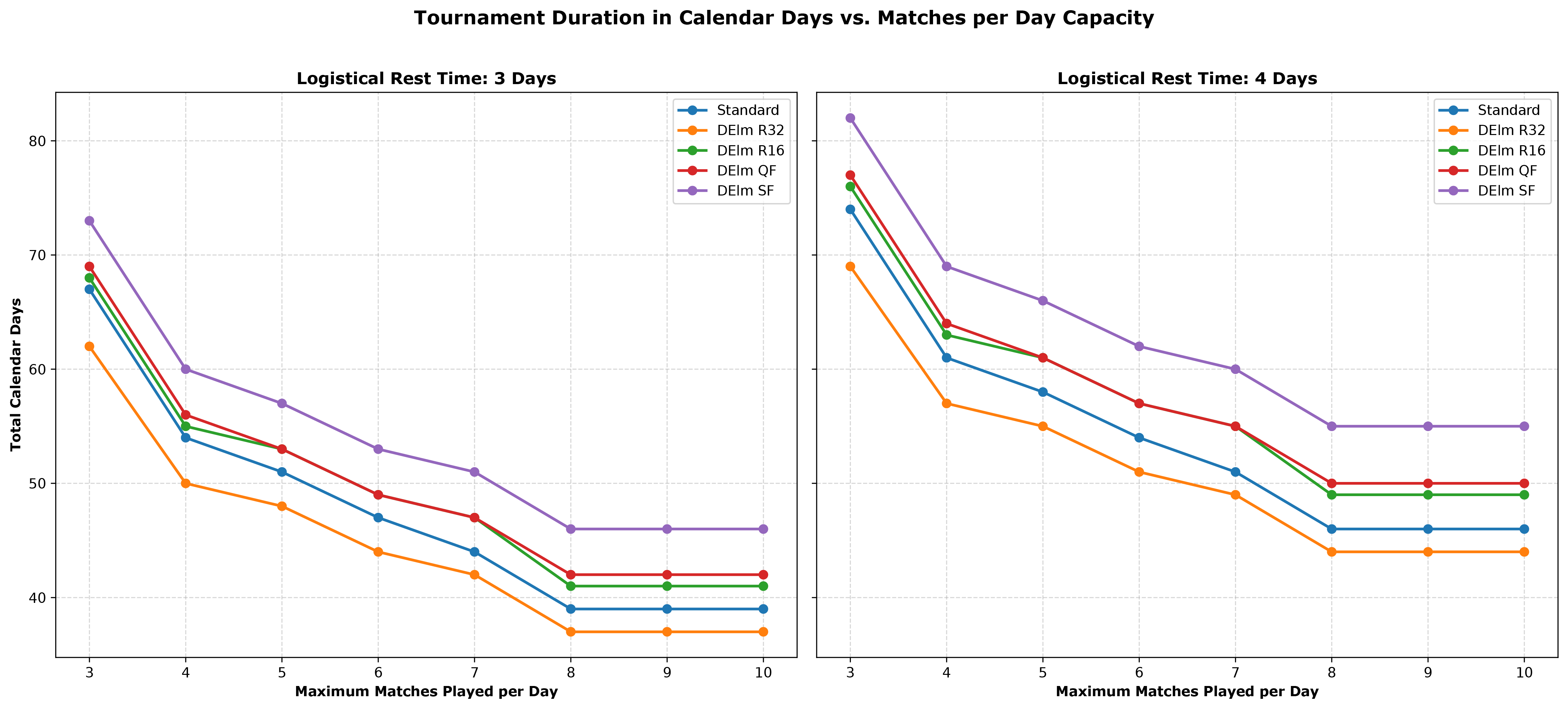}
    \caption{\textbf{Tournament Duration in Calendar Days vs. Matches per Day Capacity.} A two-panel line chart comparing the total calendar duration of the five tournament formats under a $3$-day rest period (left panel) and a $4$-day rest period (right panel). The X-axis represents the daily match broadcast capacity (ranging from $3$ to $10$ matches per day). The markers show that all formats exhibit diminishing returns as matches per day increase, with \texttt{DElm R32} representing the most compact design and \texttt{DElm SF} the longest.}
    \label{fig:scheduling_lines}
\end{figure}

To evaluate the scheduling feasibility of each format, we calculated the total calendar duration in days under different logistical constraints, varying the minimum rest period between rounds ($3$ vs. $4$ days) and the maximum matches broadcasted per day ($3$ up to $10$ matches/day) (Figure~\ref{fig:scheduling_lines}). The lines demonstrate a clear logarithmic relationship representing diminishing returns: increasing daily matches from $3$ to $5$ significantly reduces the tournament calendar (e.g., in the $3$-day rest configuration, \texttt{DElm SF} drops from $73$ to $57$ days), whereas increasing capacity from $6$ to $10$ matches/day yields very minor calendar savings. Due to its reduced number of rounds ($8$), the \texttt{DElm R32} format is consistently the shortest, requiring only $37$ days (under $3$ rest days, $8$ matches/day). Conversely, the pure double-elimination \texttt{DElm SF} format (which features $10$ rounds) is the longest, but remains highly viable, requiring just $46$ days under the same $3$-day rest and $8$ matches/day configuration. This confirms that all double-elimination formats are logistically practical for international scheduling, fitting comfortably within a $40$-to-$55$-day window when daily broadcast capacity is optimized.

\section{Discussion}

The comprehensive analysis of tournament structures for a 64-team mega-event demonstrates that the double-elimination format successfully resolves the primary structural deficiencies inherent in the traditional group stage model. Most notably, the simulations confirm that the double-elimination architecture entirely eliminates mathematically irrelevant matches, or "dead rubbers," guaranteeing that every fixture maintains absolute competitive relevance. Concurrently, this format significantly increases the frequency of high-profile games between the tournament's strongest contenders. By dynamically matching teams with similar win-loss records and providing a safety net for top-tier teams against early, high-variance upsets, the double-elimination design ensures a much higher density of elite, high-stakes matchups throughout the competition.  

These empirical results hold significant theoretical implications for tournament integrity when contextualized within the broader literature on operations research in sports. As argued by \citet{Csato2021book}, applying Operations Research is fundamental to formulating robust sports rules and improving the structural integrity of events of the FIFA World Cup's magnitude. According to the systematic review conducted by \citet{DevriesereCsatoGoossens2025}, tournament designs must be evaluated across the critical dimensions of efficacy, justice, and attractiveness. The mathematical definition of "efficacy"—specifically a system's capacity to correctly rank competitors according to their true latent strengths—is a cornerstone of modern tournament evaluation \citep{SziklaiBiroCsato2022}. In this framework, \citet{StantonWilliams2013} provide direct mathematical backing for the double-elimination format, demonstrating its structural superiority in maximizing justice and efficacy while neutralizing the random factors that often distort standard single-elimination outcomes.  

Furthermore, this structural efficacy directly translates into a robust resilience against strategic manipulation. Applying the two-dimensional manipulation taxonomy proposed by \citet{DevriesereGoossensWillem2025}, the double-elimination format effectively neutralizes opportunities for bilateral collusion. Because the tournament relies entirely on continuous knockout brackets where every match strictly dictates survival or bracket placement, teams cannot cooperatively engineer mutually beneficial outcomes, removing the vulnerability inherent in traditional round-robin group stages \citep{GuajardoKrumer2024, Csato2025collusion}.  

Despite these theoretical advantages, it is essential to candidly address the primary limitation of the double-elimination format: the inherent contradiction between mathematical efficacy and logistical planning viability. As highlighted by \citet{KendallEtAl2010}, sports scheduling encompasses immense complexity, particularly when optimizing calendars against rigid travel restrictions and venue availability. The dynamic nature of the repechage bracket introduces complex operational constraints and heightened scheduling complexity. Specifically, the unpredictable geographical paths for teams entering the losers' bracket complicate travel planning for fans and the management of visas and border crossings for tournament organizers. However, this logistical burden is partially offset by the clear predictability of the main bracket, as winning teams know their exact path to the final from their very first game. This deterministic scheduling for successful teams is arguably easier to manage than the chaotic and asymmetrical pathways generated by the current FIFA format, which relies heavily on the uncertain qualification of the best third-placed teams.  

Another critical trade-off involves the athlete workload and the highly asymmetric distribution of matches per team. While the format successfully reduces meaningless games for weaker teams, squads navigating the repechage bracket face a significant physical overload. A team suffering an early defeat might need to play up to nine or eleven matches to secure the championship. As physiologically demonstrated by \citet{LauxEtAl2015}, failing to maintain a proper recovery-stress balance significantly elevates injury risks among professional football players. This extreme match volume risks inducing severe fatigue or even strategic underperformance (tanking) as teams attempt to manage physical exhaustion. Consequently, implementing this format requires specific logistical adjustments, such as optimizing rest intervals and restricting extra time exclusively to the final stages, to protect player health and maintain match quality.  

The success of this format is also heavily dependent on second-order effects, particularly the chain of integrity originating from the qualification phases. As demonstrated by \citet{CsatoKissSzadoczki2025}, the efficacy of the final tournament is strictly conditioned by the relative quality and merit of its participants. If the expansion to 64 teams relies on arbitrary political quotas rather than empirical sporting merit, the resulting skill disparity will rapidly polarize the double-elimination bracket. This artificial disparity would inevitably lead to excessively unbalanced early matches, undermining the format's core objective of producing highly competitive fixtures.  

The expansion of the FIFA World Cup™ to 64 teams introduces profound economic pressures that extend beyond tournament logistics to the broader cultural ecosystem. The increased scale places significant financial burdens on fans through higher travel costs, extended accommodations, and inflated secondary markets. This economic strain is also evident in related cultural phenomena. For instance, \citet{AlvarengaEtAl2026} demonstrated that the tournament's expansion to a massive 980-item sticker album creates a costly collective-action dilemma. Their simulations reveal a structural parallel to tournament design: just as inflexible trading norms exacerbate financial inefficiencies and severely penalize the least fortunate collectors, rigid legacy tournament formats trap teams in irrelevant matches and punish early upsets. In both scaled systems, adopting adaptable frameworks (such as altruistic trading networks or double-elimination brackets) is essential to optimize collective efficiency and mitigate systemic costs.

Ultimately, as mega-sporting events continue to expand in scale, relying on incrementally modified versions of legacy structures is no longer sustainable. This reality is already prompting structural revolutions in other elite competitions, such as the UEFA Champions League's recent transition from traditional round-robin groups to a single-league format designed specifically to evaluate and maintain competitiveness \citep{DevriesereGoossensSpieksma2026}. In this same vein, the continuous scaling of the FIFA World Cup™ necessitates a paradigm shift toward non-traditional tournament designs to preserve competitive integrity and ensure the event remains a true test of global sporting merit.

\section{Acknowledgement}

This research was supported by the High Performance Computing Center at UFRN (NPAD/UFRN), INCT NeuroComp, CNPq, Itaú ICTi, CAPES and UFRN/PROPESQ.

\bibliographystyle{elsarticle-harv} 
\bibliography{worldcupformat}

\end{document}